# Single and binary evolution of Population III stars and their supernovae light curves

T.M. Lawlor[1], T.R. Young[2], T.A. Johnson[2] and J. MacDonald[3]
[1]Penn State University, Delaware County Campus, Department of Physics, Media, PA 19063 USA
[2]University of North Dakota, Department of Physics, Grand Forks, ND, 58202, USA
[3]University of Delaware, Department of Physics and Astronomy, Newark, DE 19716, USA

**ABSTRACT**
We present stellar evolution calculations for Population III stars for both single and binary star evolution. Our models include 10 $M_\odot$ and 16.5 $M_\odot$ single stars and a 10 $M_\odot$ model star that undergoes an episode of accretion resulting in a final mass of 16.1 $M_\odot$. For comparison, we present the evolution of a solar heavy element abundance model. We use the structure from late stage evolution models to calculate simulated supernova light curves. Light curve comparisons are made between accretion and non-accretion progenitor models, and models for single star evolution of comparable masses. Where possible, we make comparisons to previous works. Similar investigations have been carried out, but primarily for solar or near solar heavy metal abundance stars and not including both the evolution and supernovae explosions in one work.

1 INTRODUCTION

It is commonly believed that the first stars, Population III (Pop III) stars, were formed from primarily hydrogen and helium. These zero heavy element abundance objects are of great interest for a variety of reasons. They are responsible for re-ionization of the Universe and their nucleosynthetic yields are the seeds for the makeup of later stars, including our own. The cosmological effects of these primordial stars have been studied by Tumlinson, Shull & Venkatesan (2003). Recently Greif & Bromm (2006) suggested that Pop II.5 stars (intermediate between Pop II and III) are the key ingredient for enriching the early universe. No Pop III stars have been directly observed to date, which may not be surprising. They are expected to live shorter lives than today's stellar populations and they may have been polluted by the ISM.

There is still some debate on the mass spectrum, or initial mass function (IMF), of Pop III stars. It was proposed early on (Palla, Stahler & Salpeter 1983; Eryurt-Ezer 1985) that there should be a wide range of mass for Pop III stars, with a distribution that is bimodal around 1 $M_\odot$ and 100 $M_\odot$. Nakamura & Umemura (2001) also find a bimodal IMF for metal free stars, but in contrast simulations of Abel, Bryan & Norman (2000) and Bromm, Coppi & Larson (2002) do not produce low mass fragments. None the less, Omukai & Yoshii (2005) argue that many IMF determinations do not include feed back effects of massive stars already formed, which may induce lower mass fragmentation. They also cite the observed low mass, very low Z object He 0107-5240 ([Fe/H] =-5.3) as an example of a possible population III star that makes it difficult to rule out low mass fragments.

There is a growing body of evolutionary calculations for Pop III stars, though a great number of uncertainties remain (Langer 2006). Heger & Woosley (2002) and Marigo, Chiosi & Kudritzki (2003) have carried out calculations for high mass models. Stars with masses between 0.7 and 100 $M_\odot$ have been extensively studied by Marigo et al. (2001). Chieffi et al. (2001) evolved intermediate mass (4 – 8 $M_\odot$) stars from the pre-main sequence to the Asymptotic Giant Branch (AGB). Siess, Livio & Lattanzio (2002) evolved models of Pop III stars with masses between 0.8 – 20 $M_\odot$ to the AGB or, for the more massive stars, to the beginning of carbon burning. Gil-Pons et al. (2005) evolved a 9 $M_\odot$ star to the end of carbon burning and formation of a massive ONe core.

Most stars in the universe today are believed to be in binary systems (Abt & Levy 1976, 1978). If the system is a close binary, one or both stars can overflow its Roche lobe, and transfer mass to its companion through the inner Lagrangian point. Early work on the evolution of low mass interacting binaries (Kippenhahn & Weigert 1967; Lauterborn 1970) gave rise to classifying mass transfer according to the evolutionary stage of the donor at the beginning of the mass transfer phase: Case A – Main Sequence (MS), Case B – post-MS but pre-helium ignition, and Case C – post-helium core burning. Since the mass transfer rate is sensitive to whether the envelope of the donor star is radiative or convective at the beginning of the mass transfer phase, a further distinction is made between early (radiative envelope) and late (convective envelope) cases. Case A and early case B and C mass transfer is, in general, stable and proceeds more or less conservatively. Late case B and C mass transfer is usually dynamically unstable because the response of a giant star with a deep convective envelope to mass loss is to expand or, at least, to shrink more slowly than its Roche lobe (Paczyński & Sienkiewicz 1972; Podsiadlowski, Joss & Hsu 1992, hereafter PJH92). For systems with massive progenitors, close binary interaction can affect the type of supernova (SN) that will result (Chevalier 1976; PJH92). The evolution of massive close binaries has been considered by a number of authors including PJH92, Vanbeveren, De Loore & Van Rensbergen (1998), Vanbeveren (1999), and Podsiadlowski, Rappaport & Pfahl (2004). In general only systems with solar or near solar heavy element abundances have been considered.

One motivation for this work is the possibility that the first Pop III 'star' may be found by way of observing its death in a supernova explosion. In this paper we present the results of evolutionary calculations of 10 $M_\odot$ and 16.5 $M_\odot$ stars with initial total heavy element abundance $Z = 10^{-15}$ from the pre-main sequence. In addition, for the 10 $M_\odot$ star we investigate the effects of accretion from a binary companion of marginally greater mass. We evolve a solar heavy element abundance star for comparison. For each evolved Pop III star, we use the structure from late stage evolution as the starting point for simulation of their supernova light curves. In Section 2 we briefly describe our stellar evolution and radiation-hydrodynamics codes. In Section 3 we present evolution results and in Section 4, we calculate and discuss supernova light curve simulations. Conclusions and discussion are in Section 5.

## 2 STELLAR CODES

2.1 The stellar evolution program

Our stellar evolution code is described in detail in Lawlor & MacDonald (2006). Here we describe the updates relevant to the current work. The number of composition variables followed in detail has been increased to H, $^3$He, $^4$He, $^{12}$C, $^{13}$C, $^{14}$N, $^{16}$O, $^{20}$Ne, $^{24}$Mg, $^{28}$Si and $^{56}$Ni. The α-chain reaction rates are taken from Timmes, Hoffman & Woosley (2000). Neutrino loss rates are from Itoh et al. (1996). Production of electron-positron pairs is now included in the equation of state.

For single stars, mass loss is treated by using a Reimers' mass loss formula (with mass loss parameter $\eta = 0.38$) for cool stars, and a modified form of the line-driven wind mass loss rate of Abbott (1982) for hot stars. For the comparison Pop I model, additional mass loss on the AGB is included by fitting the observed mass loss rates of Mira variables and OH/IR stars to their pulsation periods. Further details of the mass loss treatment can be found in Lawlor & MacDonald (2006). For Pop III stars, mass loss from hot star winds is essentially negligible.

For Roche lobe filling stars, we use a treatment similar to that of Ritter (1988). We assume that the envelope structure is isothermal above the photosphere and polytropic below the photosphere. We further assume that the envelope is in hydrostatic equilibrium in the direction orthogonal to equipotential surfaces, and that matter outside the star's Roche lobe of radius $R_L$, flows on equipotential surfaces at the sound speed. With the assumption that the scale height at the photosphere, $H_0$, is much smaller than the stellar radius, $R$, the mass loss rate due to Roche lobe overflow is then

$$\dot{M} = \begin{cases} -4\rho_0 H_0 R_L c_0 \exp\left(\dfrac{R-R_L}{H_0}\right) & \text{for } R \leq R_L, \\ -4\rho_0 H_0 R_L c_0 \left[\dfrac{2n+2}{2n+3}\left(1+\dfrac{1}{n+1}\dfrac{R-R_L}{H_0}\right)^{n+\frac{3}{2}} + \dfrac{1}{2n+3}\right] & \text{for } R \geq R_L, \end{cases} \quad (1.1)$$

where $\rho_0$, $c_0$ and $n$ are the density, sound speed and polytropic index at the photosphere.

2.2 Stellar explosion code

The radiation hydrodynamics code used for supernova light curve calculations is described in detail in Young (2004), but we describe it briefly here. The code used for SN simulations is a one dimensional Lagrangian radiation-hydrodynamics code (Sutherland & Wheeler, 1984). The input models for the code come from our evolution models at an evolutionary point near core collapse. The code assumes spherical symmetry and that the neutron star's mass is set as an inner gravitational boundary. The simulation is initiated by artificially placing thermal and kinetic energy in the inner zones, extracted from the formation of the neutron star. The injection of energy in the inner zones produces a density discontinuity that then propagates as an outward moving shock wave. The shock takes several hours to exit the star. The simulation follows the progression of the supernova material as it expands and releases the energy deposited by the shock. In all of the models shown here, only the thermal release of the shocked material is powering the light curve. In future work we will extend the detailed simulations to include radioactive ejecta. Once this is included, the simulated SN light curve remains bright for 400 days due to the release of gamma-rays from radioactive $^{56}$Co (Young 2004) and is seen in SN 1987A (Catchpole et al. 1987). The amount of mixing of radioactive material is assumed to extend through the He rich layer of the star produced during its evolution. Our simulations follow the photosphere as it recedes into the expanding material and a resultant light curve is obtained.

One modification to the code is the use of the Rosseland mean opacity tables acquired from the OPAL Opacity group at Lawrence Livermore National Laboratory: http://phys.llnl.gov/Research/OPAL/opal.html. An interpolation program provided by the OPAL group integrated into the supernova simulation code BOOM interpolates opacity values.

## 3. STELLAR EVOLUTION RESULTS

3.1 Single star evolution models

We follow single star evolution for Pop III, 10 M$_\odot$ and 16.5 M$_\odot$ models of initial composition $X = 0.765$, $Y = 0.235$, and $Z = 10^{-15}$, and, for comparison purposes, a Pop I 10 M$_\odot$ model of composition $X = 0.705$, $Y = 0.275$, and $Z = 0.020$. In Fig. 1 we compare the evolution tracks of the 10 M$_\odot$, Pop III model with those of the 16.5 M$_\odot$, Pop III model and the 10 M$_\odot$, Pop I model, all from pre-main sequence. As expected our 10 M$_\odot$, Pop III model is systematically more luminous than the Pop I model of same mass, and evolves more quickly (Table 1). For our 10 M$_\odot$, Pop III and Pop I models only, we include composition variables up to $^{56}$Ni and we include α-chain reactions. For all other calculations, only the composition equations for H, $^3$He, $^4$He, $^{12}$C, $^{14}$N, $^{16}$O and $^{24}$Mg are solved, and we include only reactions up to carbon burning. We find that the evolutionary differences from using the two networks are small for phases before neon burning.

Up to the end of core helium burning, our 16.5 M$_\odot$, Pop III model is more luminous than the 10 M$_\odot$, Pop III model. During the shell He burning phase of the 16.5 M$_\odot$ model in which the star is expanding and evolving to the red, the surface convection zone grows inward and takes H to high temperature regions where $^{12}$C has been produced by helium-burning. The resulting hydrogen-shell flash causes rapid expansion, significant dimming and extinguishes the helium burning shell. This phase lasts for about $10^3$ yr, after which the star returns to its previous luminosity but at much larger radius and with shell hydrogen burning the dominant energy source. At this point, the helium-exhausted core has a mass of 2.53 M$_\odot$. We illustrate these structural changes as a function of time in Fig. 2. The diminishing luminosity is accompanied by a decrease in effective temperature by nearly 20,000K and an increase in radius. Evidence of this late hydrogen-shell flash is illustrated in the left panel of Fig. 2, which shows the central temperature, density, and total mass inside the hydrogen shell (M$_{IHS}$) also as a function of time. M$_{IHS}$ decreases around 1.04 $10^7$ yr due to the growing convection zone, shown in Fig. 3 beginning around that time ($t = 1.038\ 10^7$ yr). The hydrogen-shell flash causes the convection zone to grow outwards toward the surface and mixes the products of helium burning and subsequent

CNO cycling throughout most of the envelope. This increase in heavy element abundance results in higher opacity in the envelope and a larger star. However, the convection zone does not reach the photosphere and the surface abundances are unchanged at this point. Carbon burning begins at the centre and continues for 3,800 yr, during which time the innermost 0.75 $M_\odot$ is depleted of carbon. Carbon burning then proceeds by shell burning. At the beginning of the shell burning, an existing thin surface convection zone grows deeper and dredges nuclear-processed material to the photospheric regions. This shell burning phase lasts about 600 yr and by the end, the innermost 1.2 $M_\odot$ are mostly depleted of carbon. A second shell burning episode quickly ensues. The convection zone above the burning shell grows outward and reaches the tail of the helium profile left from prior shell helium burning. A small amount of helium is ingested. The rapid energy release from helium burning reactions causes rapid further growth of the convection zone and ingestion of more helium. This leads to a large increase in the rate of energy production from helium burning reactions, with a peak rate of $2.5 \; 10^{11}$ $L_\odot$. Expansion and cooling of overlying and underlying layers halt the hydrogen and carbon burning, and the star enters a helium burning phase that lasts for 2,500 yr. During this phase the carbon burning shell is re-established, and the convection zone above the carbon burning shell merges with the outer convection zone so that now hydrogen and helium are both mixed to high temperature regions. A second hydrogen flash occurs that extinguishes helium and carbon burning. After the flash, the star settles into a phase of burning hydrogen and helium quiescently in shells that lasts for about 25,000 yr, before the helium shell becomes thermally unstable. The helium exhausted core has a mass of 1.29 $M_\odot$ at the beginning of this phase and grows due to shell burning to 1.33 $M_\odot$ at the end of this phase. Due to Reimers' law mass loss, the star at this point has been reduced to 11.7 $M_\odot$. The mass loss rate, $\sim 2 \; 10^{-6}$ $M_\odot$ yr$^{-1}$, is about thrice the rate at which the core is growing in mass. Hence unless there is a significant increase in mass loss rate during the thermally pulsing AGB phase the core mass will eventually exceed the Chandrasekhar mass and core collapse will occur. The locations of the boundaries of convection zones and the hydrogen and helium-exhausted regions are shown in Fig. 3 for the 16.5 $M_\odot$, Pop III model from near the end of core helium burning forward to end of the calculation.

The evolutionary phases up to the hydrogen shell flash of our Pop III, 10 $M_\odot$ model are similar to those of the 9 $M_\odot$ model of Gil-Pons et al. (2005). However the subsequent evolution differs significantly in a number of ways. The initial hydrogen flash is followed by a series of smaller hydrogen flashes as the convection zone continues to grow inwards. Because a large amount of computer time is needed to follow each of these flashes, we suppress them after the 15$^{th}$ flash by reducing the convective mixing efficiency from $\beta = 0.1$ to $\beta = 10^{-4}$. In contrast to Gil-Pons et al. (2005) we do not find that thermal pulses occur. Instead hydrogen and helium shell burning proceed quiescently with an inter-shell mass of about 10$^{-5}$ $M_\odot$. As a result, a carbon layer builds up below the helium shell. After 40,000 yrs of double shell burning during which the core mass increases from 1.315 to 1.388 $M_\odot$, the carbon ignites under moderately degenerate conditions. During the ensuing carbon shell flash, the peak nuclear energy generation rate exceeds $5 \; 10^{15}$ erg gm$^{-1}$ s$^{-1}$, with the total energy generation rate exceeding $3 \; 10^{13}$ $L_\odot$. At the point where we end the calculation, velocities in the envelope exceed 40% of the local escape velocity and hydrodynamical effects are becoming important. Flash driven convection has reached the helium layer and helium is being mixed towards regions where the temperature has reached $2 \times 10^9$ K. Hence it is possible that the envelope is ejected before core collapse occurs.

The evolution of our Pop I, 10 M$_\odot$ model is similar to that found by Ritossa, García-Berro & Iben (1996) and Iben, Ritossa & García-Berro (1997) in their studies of 10.0 M$_\odot$ and 10.5 M$_\odot$ Pop I stars. One important difference is that our model does not experience a blue loop during core helium burning. A second difference is that the mass of the He-exhausted core at the end of core carbon burning, 1.283 M$_\odot$, is significantly higher than the 1.206 M$_\odot$ found by Ritossa et al. (1996), and is closer to the value, 1.269 M$_\odot$ found by Iben et al. (1997) for their 10.5 M$_\odot$ Pop I star. At the beginning of the thermally pulsing super AGB phase, the core mass has increased to 1.291 M$_\odot$. We have followed the first 270 thermal pulses, which take place over an interval of ~ 7,000 yr. During this time the core increases in mass by 0.0065 M$_\odot$ and the stellar mass decreases by 0.9 M$_\odot$. Thus the evolution is dominated by mass loss. Extrapolation of these results indicates that the envelope will be totally removed when the core has mass ~ 1.34 M$_\odot$, and hence the star will likely end as a massive O-Ne white dwarf. If mass loss is overestimated, it is possible that this model will result in a core collapse SN (if the resulting core is greater than 1.35 M$_\odot$).

In Table 1, we compare the time-scales for the duration of the core hydrogen, helium and carbon burning phases for our three models with other values in the literature. Our hydrogen and helium core burning durations are in quite good agreement with previous evolution calculations.

We show in Table 2 selected physical characteristics for our two Pop III models. Because the Pop III opacities are much lower than for Pop I or II abundances, the stellar radii of our two Pop III models are significantly smaller than Pop I or II stars of the same mass and evolutionary state, until the episode of significant dredge up of heavy elements that occurs shortly after the end of core carbon burning. As a consequence of the small radii, mass loss is essentially negligible before dredge up.

The surface chemical abundances for H, He, C, N, O, and Mg at selected evolutionary phases are given in Table 3. At the time of deepest extent of the surface convection zone in their 9 M$_\odot$ sequence, Gil-Pons et al. (2005) find envelope abundances $X(^{12}C)=2.04 \times 10^{-4}$, $X(^{14}N) = 2.41 \times 10^{-6}$ and $X(^{16}O) = 3.47 \times 10^{-6}$. In comparison, we find for our 10 M$_\odot$ sequence at the equivalent stage, $X(^{12}C)=2.93 \times 10^{-4}$, $X(^{14}N) = 3.33 \times 10^{-6}$ and $X(^{16}O) = 7.82 \times 10^{-5}$. The carbon and nitrogen abundances are similar but our oxygen abundance is considerably higher. Similar to the lower mass model, the surface abundances for our 16.5 M$_\odot$ model remain essentially constant until a point after the onset of carbon burning. In this case, however there is a much more significant enrichment of the envelope with metals. Most significantly, the nitrogen mass fraction is orders of magnitude higher than in the 10 M$_\odot$ model. This may be pertinent to unusually high nitrogen abundances found in the extremely metal poor object CS 22949-037. This object was found to have an extreme overabundance of nitrogen relative to iron (Norris, Ryan & Beers 2001, here after NRB01). Norris et al. (2002) have suggested that this and other unusual abundances in this star are associated with element production in rotating massive (> 200 M$_\odot$) Pop III stars that end their lives in hypernova explosions. Alternatively, as indicated here, less massive stars may be responsible for the overabundance of nitrogen.

## 3.2 Pop III, 10 $M_\odot$ model with accretion

We consider a binary system in which the stars have similar mass of about 10 $M_\odot$. We choose an initial separation so as to have Roche lobe overflow occur when the slightly more massive primary is undergoing core helium burning. The envelope is mainly radiative and hence mass transfer is stable early type C. We investigate this particular scenario because the core helium burning phase has a longer lifetime than the other post MS phases, and also to avoid the common envelope phase of evolution that is likely to develop from late B or C mass transfer in which dynamical rates as high as 100 $M_\odot$ yr$^{-1}$ can occur.

For a Pop III, 10 $M_\odot$ primary with an initial binary separation of 6 10$^{11}$ cm, Roche lobe overflow occurs when the primary radius is 2.3 10$^{11}$ cm. We assume that mass transfer is conservative and hence the rate of accretion onto the secondary is equal to the mass transfer rate. Because very little processed material is dredged-up to the surface until after core carbon burning, the accreted matter is given the same composition as the envelope of the secondary.

Here we focus on the evolution of the mass receiving secondary. This model is a Pop III, 10 $M_\odot$ binary progenitor, evolved including an episode of accretion, described in Section 2.1. For consistency and brevity, and because we only include accretion in one model, we will here after refer to this model as our 'accretion model.' Similar calculations were carried out for solar heavy element abundance models by PJH92 for accretion occurring before and after the end of core hydrogen burning. In that paper, they determine a fractional core mass criterion($\xi_c = m_c/M$) for the appearance of the final pre-supernova model being either a blue or red supergiant. In this definition, $m_c$ is the hydrogen depleted core mass and $M$ is the total stellar mass. We do not attempt to determine such a criterion; rather we only describe evolution following accretion so that we can calculate its subsequent supernova light curve in Section 4. The shape of the evolution track for our accretion model more closely resembles PJH92's models for which accretion occurs during the main sequence. This is probably because we begin accretion a short time after the onset of core helium burning, thus there is evidently some transition region distinguishing their two sets of accretion models or a fundamental difference for stars of low and high Z. We defer a larger grid of models as a function of the accretion turn-on point to a later paper and focus here on the ensuing supernovae for select models.

Accretion significantly changes the evolution of our 10 $M_\odot$ Pop III progenitor model (Fig. 4). It experiences an increase in luminosity and temperature during the period of mass transfer on a time scale of 10$^5$ – 10$^6$ years. During this time Log (L/L$_\odot$) increases from 4.25 to 4.8 and Log(T$_{eff}$) from 4.63 to 4.73. Considerable dimming following accretion lasts for 10$^4$ years, however the subsequent increase in luminosity to Log (L/L$_\odot$) = 4.5 occurs in only ~300 years. The entire rise in luminosity occurs on the order of 10$^3$ years, so although this evolution looks very similar to the single star evolution (Fig. 4), the time-scale is longer. The loop near Log (L/L$_\odot$) = 4.7 lasts for 15,000 years and the model moves to the red in 5.0 10$^5$ years, which is roughly twice as fast as the single star's evolution. This may not be surprising since the model is now more massive, reaching 16.1 $M_\odot$. The prior delay in evolution to the red is consistent with the notion (for Pop I stars) that a decreasing fractional core mass due to accretion prevents evolution to the red. None the less, our Pop III models all end up as red supergiants. Time-scales are illustrated in Fig. 5 and 6. In Fig. 5 and in the right panel of Fig. 6, the effects of accretion induced hydrogen pulses are shown. We attribute these hydrogen pulses to compressional heating due to increased episodes of accretion and to convective mixing.

We find in Fig. 5 that central stellar conditions for accretion model are significantly changed compared with a single star model of equal progenitor mass and heavy metal abundance. Accretion delays carbon burning and thus core collapse. Carbon burning luminosity initially turns on following the end of accretion (point cI in Fig. 4), but stalls until late hydrogen pulses cease, near point cII in Fig. 4. For this model the stellar lifetime is increased by ~5%, even though the final mass is greater. For single star evolution, our values for central temperature and density are quite comparable to the values of Siess et al. (2002) for their models of the same mass, and at the same phases of evolution. They report central temperature and density at a point during hydrogen and helium burning when central abundances for $X$ and $Y$ (respectively) have dropped to 0.50. For their 10 $M_\odot$ model during hydrogen burning, they report $T_c = 9.17 \ 10^7$ K and log $\rho_c$ = 2.398, while for our models $T_c = 9.12 \ 10^7$ K and log $\rho_c = 2.389$. We also find good agreement with their values during helium burning. They find $T_c = 1.699 \ 10^8$ K and log $\rho_c = 3.132$ and our values at the same point are $T_c = 1.71 \ 10^8$ K and log $\rho_c = 3.204$. No values for a Pop III binary model of this specific mass have been published.

In the left panel of Fig. 5 the evolution of H, He, and C nuclear burning luminosities are shown. The top panel shows these for our original Pop III 10 $M_\odot$ model and the bottom panel shows the same luminosities our accretion model. Accreted mass induces a number of hydrogen pulses. Similar to what we found for our single 16.5 $M_\odot$ model, the first hydrogen pulse shown in Fig. 5 is accompanied by a growing convection zone, shown in Fig. 7, mixing down a fresh supply of protons. During each subsequent H-pulse the model star swells in size, decreasing density and damping out both helium and carbon burning temporarily. Initially carbon core burning begins at the same time as for the single star model, but due to accretion induced hydrogen pulses, it is delayed until a later point, thus extending the lifetime of the star slightly.

The final surface abundances of X and Y for accretion and non-accretion models are comparable but there is a market different in nitrogen abundance (Table 4). For our accretion model, $X(^{14}N) = 6.46 \ 10^{-3}$ while the non-accretion model of the same progenitor mass has $X(^{14}N) = 5.5 \ 10^{-6}$. This happens by first making C by the triple alpha process and then mixing in of protons to make N. Further, C/O is on the order of 10 for the non-accretion model, while the accretion model has only C/O = 0.1. In the later case, more oxygen is dredged up. Our accretion model is fully convective during late stage evolution (Fig. 7) which is as for all three Pop III models, the only period for which we find any significant dredge up to the surface. This is consistent with what Siess et al. (2002) found for single star models with mass greater than 15 $M_\odot$ and with models of Marigo et al. (2001).

In the following section we present light curve calculations for each of our final evolution models and make comparisons between light curves from Pop III progenitors and between accretion and non-accretion pre-supernova models.

# 4 POP III SUPERNOVAE SIMULATIONS

As described above, evolution calculations used in the SN code proceed as far as carbon burning, which produces various mass oxygen cores. It is expected an iron core will develop later but on a short time scales compared to earlier burning cycles (Arnett 1994). Our future calculations will include evolution models with fusion processes beyond carbon and for a wider range of masses. It should be presumed that all light curves presented in this section are calculated from progenitor evolution models that are Pop III, thus we suppress this label when referring to evolution models. For each evolution model we calculate two light curves using two metallicities, Z, in the stellar explosion code. By doing this, we illustrate a range which depends on the extent of pollution in the photosphere when the star explodes. For this purpose, we use $Z = 0.0$ and $Z = 0.008$. Thus upper bound of this metallicity range is near what we find for the final Z in our accretion model.

The outcomes of the evolution models play two roles in understanding the supernova explosion. First, the mass and radius of the star determine the shock wave deposition of energy and the early radiating surface area. Second, the stellar composition determines the rate at which the photosphere recedes into the material. Figures 8 -10 show the composition versus enclosed mass of three models at the end of the evolution calculations. Figures 8 and 9 are for our single Pop III models and Fig. 10 is for our accretion model. Fig. 8 shows the 10 $M_\odot$ model ending with a 2 $M_\odot$ He core and an 8 $M_\odot$ H envelope. Fig. 9 shows the 16.5 $M_\odot$ model with a large O core of about 13 $M_\odot$ and a smaller 3 $M_\odot$ H envelope. In Fig. 10 our accretion model is shown to evolve a larger He core (7 $M_\odot$) as compared to the single star. The core collapse energy in forming a neutron star of mass 1.35 to 1.38 $M_\odot$ (Thorsett & Chakrabarty 1999) is set as a thermal energy source to initiate the explosion above the inner boundary condition. This explosion energy quickly develops into an outward travelling shock wave that disrupts the star.

Table 5 shows five important parameters of the three pre-supernova stars used in the simulated explosions. In general a light curve simulation depends on these parameters: progenitor radius, ejected mass, explosion energy, Ni mass, and Ni Mixing. All Pop III stellar explosions simulated have kinetic energies close to $1 \times 10^{51}$ ergs, which are typical of core collapse SNe.

In future calculations radioactive $^{56}$Ni ejecta will be included as an energy source that powers the late time light curve. This can be particularly important if the radius is small, as was the case with SN 1987A (Arnett et al. 1989). We also plan to and examine explosion simulations for the donor star in which Ni most likely will power the light curve entirely. The three important parameters that change the resulting light curve considered here then become the stellar radii, mass of the evolved stars, and the mixing of Ni. We assume the mixing of Ni extends through the He core due to Rayleigh-Taylor instabilities. The extent of mixing changes with each these models due to the variation in helium core mass. The mixing will extend to different mass positions and change the nature of the resulting light curve. In the code used here, these are treated using simple assumptions that the composition boundaries C/O/He will provide the necessary gradients in an accelerated hydrodynamic flow. A further assumption is that explosive nucleosynthesis is not included. Once the shock reaches the surface the star, it leaves the stellar gas in a super heated state and expanding behind the accelerating shock wave. The photosphere at this point is close to the outer layers of the star and moving outward. This period in the evolution of the explosion usually corresponds to the brightest part of the light curve. This is illustrated in Fig. 11 in which the accretion model shows the largest peak luminosity and has the greatest radius (see Table 5). Where the brightest part of the light curve occurs does however depend dramatically on the initial radius and can be much later if the initial radius is small. As the material expands, the photosphere recedes due to expansion and the opacity drops. This is particularly important for stars that lack metals. It is important to note that the dominate opacity is electron scattering and occurs for all easily ionized atmospheres, typically H. Metals in the atmosphere are expected to change the opacity due to many excitation lines, but not as much an effect was seen in the model with the most H in the envelope. Fig. 12 shows the light curves of two models each with $Z = 0.0$ and $Z = 0.008$. In the accretion model the difference in brightness is small. However in the single 16.5 $M_\odot$ model, the differences are much more pronounced, with the luminosity peak 2.5 magnitudes brighter and the plateau reduced by about 30 days for the $Z = 0.0$ case. The simplest explanation for this is the large oxygen core of the single 16.5 $M_\odot$ star results in a higher recombination temperature that creates a faster light curve evolution. Thus the movement of photosphere inward is much more rapid with low abundances of other metals. This is also seen in the velocity profiles (Fig. 13) when comparing the 16.5 $M_\odot$ single star model light curves with different metallicities, $Z = 0.0$ and the $Z = 0.008$. The velocity of the 16.5 $M_\odot$ single model with $Z = 0.0$ falls more rapidly inward to lower velocities at any given time past about 15 days. Further inspection of the light curve in Fig. 12 concerning the accretion model shows the same trend of early release of energy and a shorter plateau occurring for the $Z = 0.0$ but on a smaller scale. This identifies a possible trend occurring with Pop III stars producing a fainter peak and longer plateau. In our simulations there are no non-thermal contributions to the opacity except for an opacity floor which is set at 0.025 for He layers and 0.001 for hydrogen. Changing these with respect to the models metal abundance will be further explored in future work. We used the same opacity floors for all models considered here. This also presents an interesting possibility that when Ni is included in the simulations that it will have a more dramatic effect on the light curve than starting with higher metallicities. For $Z = 0.0$ models the photosphere is reaching inner material that has been heated by the radioactivity faster than when metals are present. The plateau will become brighter and stay bright for a longer time than the higher metal content supernovae. Thus the energy that is produced by the radioactivity has a larger impact on the light curve than previously suspected. This is because most of the radioactivity heating, over time, gets absorbed in the PdV expansion. The photosphere is still in the outer envelope of the ejecta and will not see the radioactive contribution. In low metal stars the radioactivity heating wave is revealed earlier in the evolution of the light curve.

## 5 CONCLUSIONS AND DISCUSION

We have presented Pop III evolution calculations for two masses and for one that undergoes mass transfer. In each case we found to some degree, a series of late hydrogen pulses. Pulses are most pronounced for models that include accretion, which can be partly attributed to compressional heating, but also due to deep convective mixing. One secondary effect of accretion is a modest increase in the lifetime of the star. The final surface abundances for a model including accretion are modestly different than our 16.5 $M_\odot$ single star model, but overall on the same order. The modest differences include slightly less helium and oxygen, and slightly more nitrogen. The accretion model does end up with an order of magnitude less carbon and other heavier metals, and it also has a higher hydrogen surface abundance, but this is most likely due to accreting matter from an unpolluted companion. The differences between our accretion model and 10 $M_\odot$ single star model are significant, as was the case when comparing our 16.5 $M_\odot$ model. At least during pre-SN evolution and following the end of accretion, our accretion model evolution resembles a 16.5 $M_\odot$ single star model when comparing evolution tracks. However, the time-scale of the accretion model evolution is dissimilar then that of the 16.5 $M_\odot$, primarily due to induced hydrogen pulses. Interestingly the light curve for the accretion model is brighter than the single star model even though the final mass of the accretion model is 0.5 $M_\odot$ less. This is due to the larger radius of the accretion model. With a larger radius the shock wave spends more time in the presupernova star and less energy goes into the expansion of the material. This allows a brighter peak and plateau phase of the supernova light curve. The model did have an artificial excess in kinetic energy, about 40% needed to initiate a shock, which accounts for some of the increased luminosity but the radius has been shown to be a more influential parameter (Young 2004). In addition the excess energy would have produced a shorter plateau and this is not seen in the simulations suggesting a greater influence from the progenitor radius.

We have found that our single 16.5 $M_\odot$ Pop III model experiences significantly more dredge up of nitrogen compared to our single 10.0 $M_\odot$ Pop III model. This implies that an over abundance of nitrogen may be possible to explain with relatively low mass stars. The resulting supernova light curve had a secondary peak in the plateau even without Ni present in the calculation. This can be attributed to the unusual large core of the presupernova star. The rapid cooling of the core is very evident by the velocity profile. The explanation for this is that the stored thermal energy is being released rapidly because the opacity is low. This energy is seen as the secondary peak in the light curve.

Our single 10 $M_\odot$, Pop III model experienced a late carbon shell flash, which may result in ejecting the envelope prior to core collapse. This effect has not been included in supernova simulations presented here. Our comparison 10 $M_\odot$, solar heavy metal abundance model, evolved with the absence of a blue loop during helium burning and resulted in a higher core mass compared with previous works. We estimate a final core mass ~1.34 $M_\odot$, which is at the limit for a core collapse supernova.

Based on our initial small sample of models, we identify a possible trend that Pop III SN show a fainter peak and longer plateau. Our future goal is to produce an extensive Pop III evolution model grid completely through their stellar explosions, which may be available for comparison to potential candidates for Pop III supernovae. Similar modeling investigations have been carried out previously (by PJH92, for example) but neither for metal-free stars nor the resulting supernova simulations.

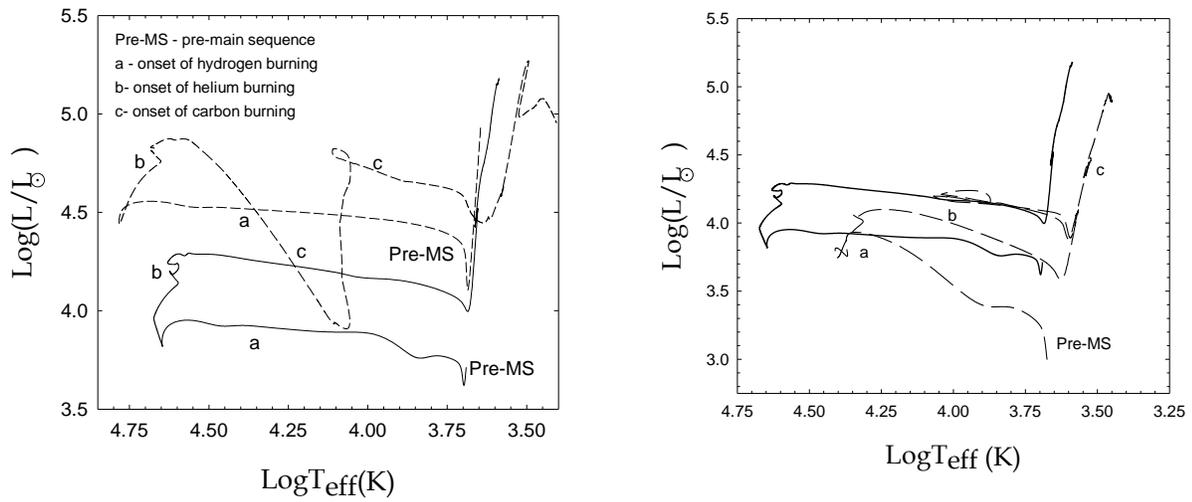

Figure 1 Hertzsprung-Russell diagrams. *Left:* Zero heavy element abundance models for 10 M$_\odot$ (solid line) and 16.5 M$_\odot$ (dashed line). *Right:* Zero heavy element abundance (solid line) and solar heavy element abundance (dashed line) models both for 10 M$_\odot$.

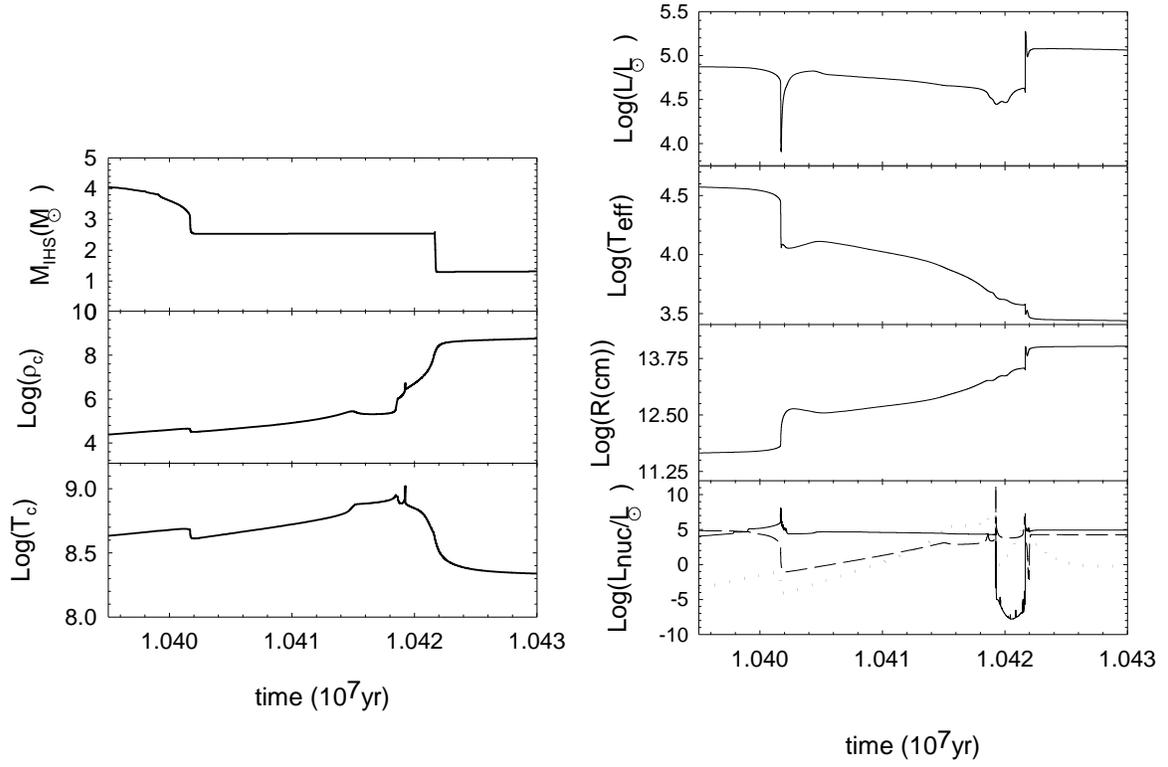

Figure 2 *Left:* Mass interior to the hydrogen shell ($M_{IHS}$) (top) and the central density and temperature for a 16.5 $M_\odot$, Pop III model. *Right:* Luminosity, temperature, radius and nuclear burning luminosity for the same model. In the bottom panel we include hydrogen (solid line), helium (dashed), and carbon (dotted) burning luminosities

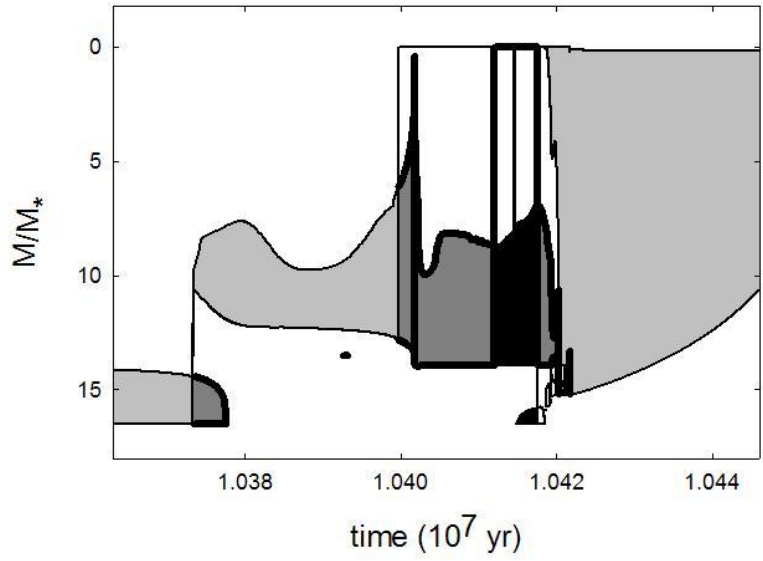

Figure 3. Convection zones for a 16.5 M$_\odot$, Pop III model. The vertical axis is mass above the convective boundaries, thus the surface is at zero. The total mass of the star is essentially 16.5 M$_\odot$ until after t ~ 1.042 10$^7$yr. The dark grey episode of convection shown near 1.040 10$^7$yr induces the late hydrogen pulse shown in the right panel of Fig. 2. The final deep convective zone (light grey) is responsible for dredge up to the surface.

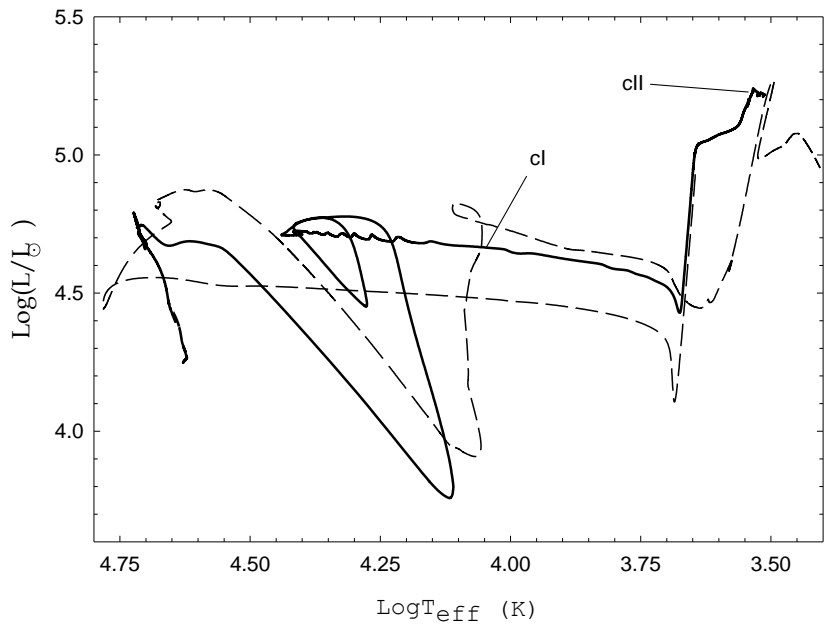

Figure 4. HR diagram for the evolution of our accretion model with final mass 16.1 $M_\odot$, and for a Pop III, 16.5 $M_\odot$ model (dashed line). Label cI is the initial more modest onset of carbon burning and cII is the point where carbon burning eventually turns on (see Fig. 5). Accretion is turned on at the beginning of the accretion model's evolution shown here, and corresponds to a time shortly after the onset of core helium burning.

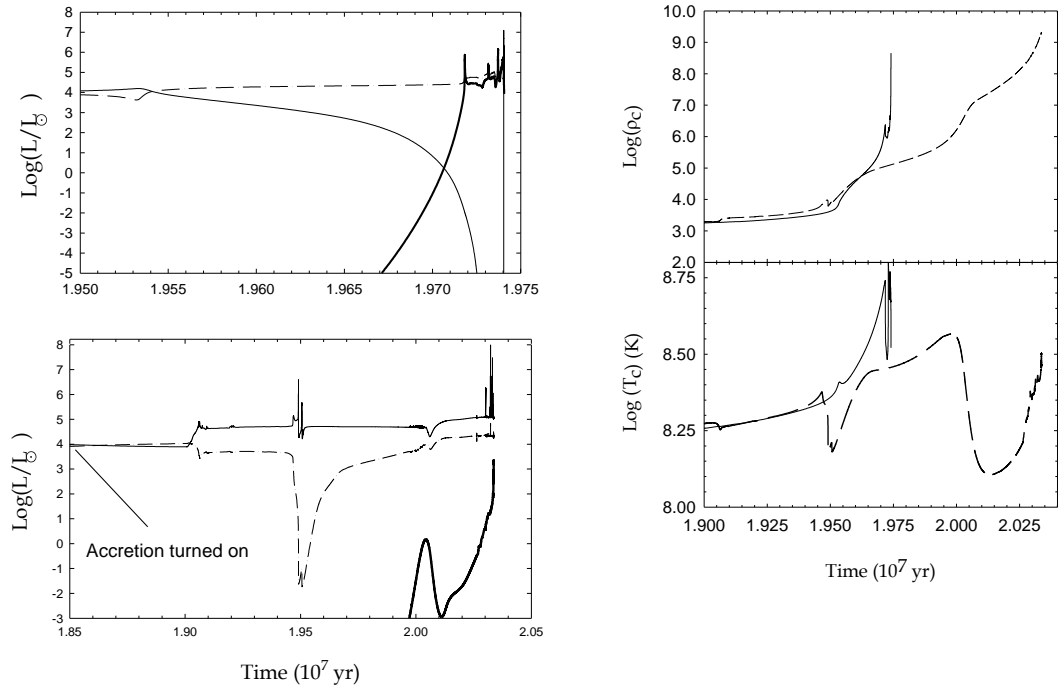

Figure 5. Left: The hydrogen (solid), helium (dashed), and carbon (thick solid) burning luminosities for a 10 M$_\odot$, Pop III model (top) and for our accretion model bottom). Right: The central density and central temperature for the same models (dashed line = accretion model).

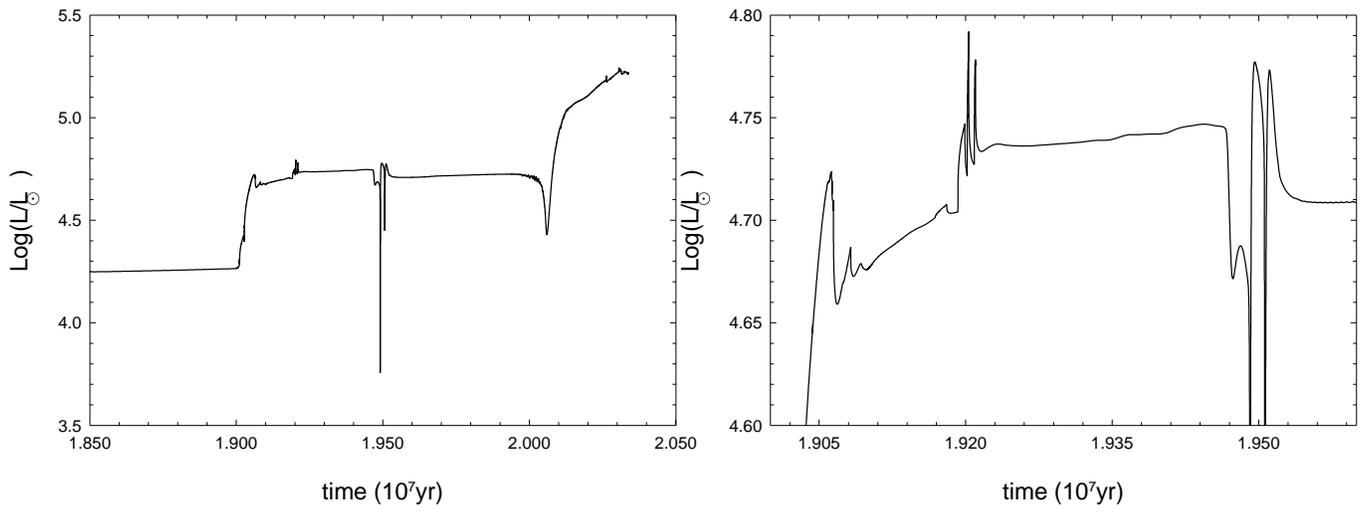

Figure 6. Left: Luminosity vs. time for an accretion model. Right: The same, but an expanded view of the effect of accretion induced hydrogen pulses.

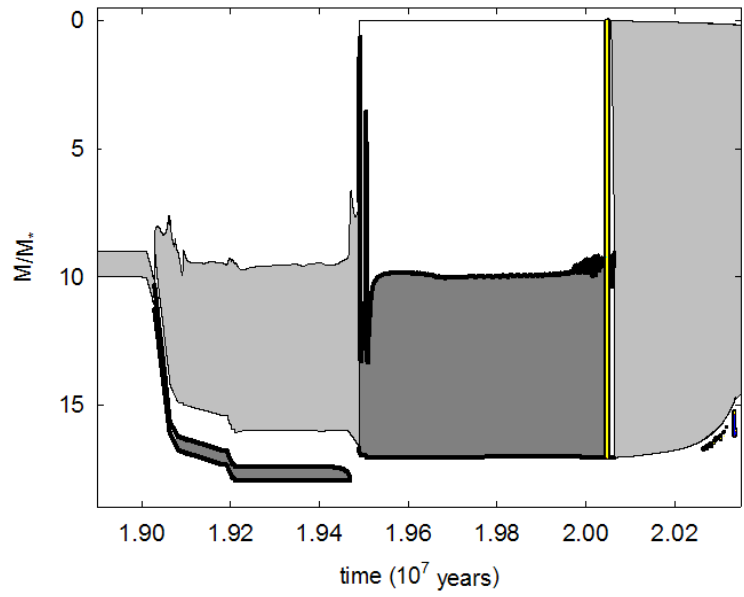

Figure 7. Evolution of convective shells for our accretion model. The vertical axis represents the mass above each boundary of convection regions, and the stellar surface is at zero.

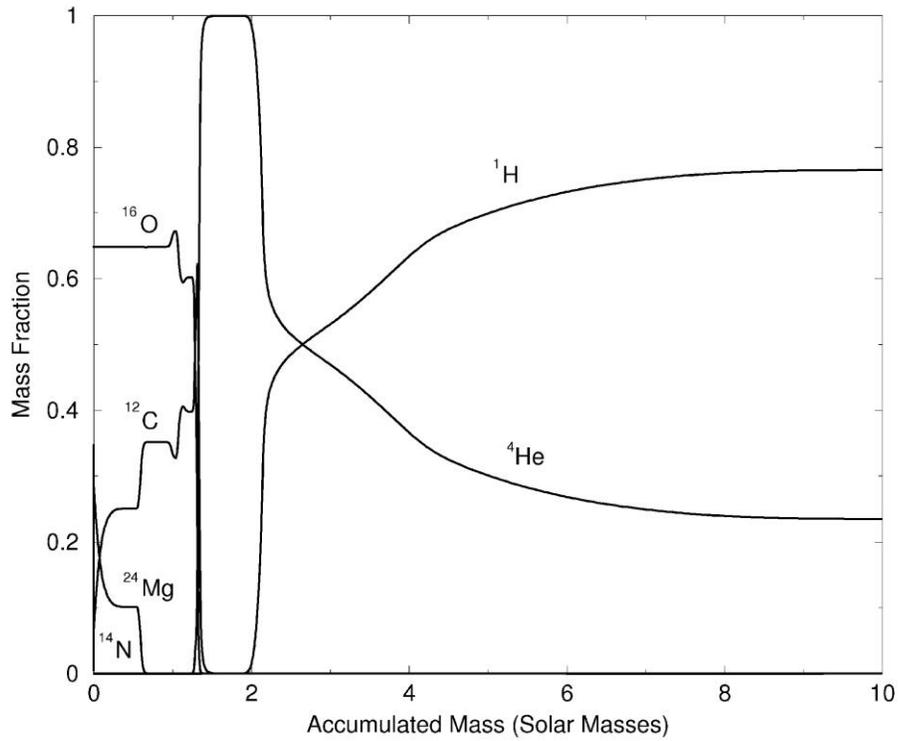

Figure 8. Mass Fractions (as labelled) against mass coordinate for the Pop III, 10 $M_\odot$ single star model. The oxygen/carbon layer is about 1.5 $M_\odot$, the He layer is 0.5 $M_\odot$ and the H layer is 8 $M_\odot$. In this model, $^{24}$Mg represents all elements heavier than carbon.

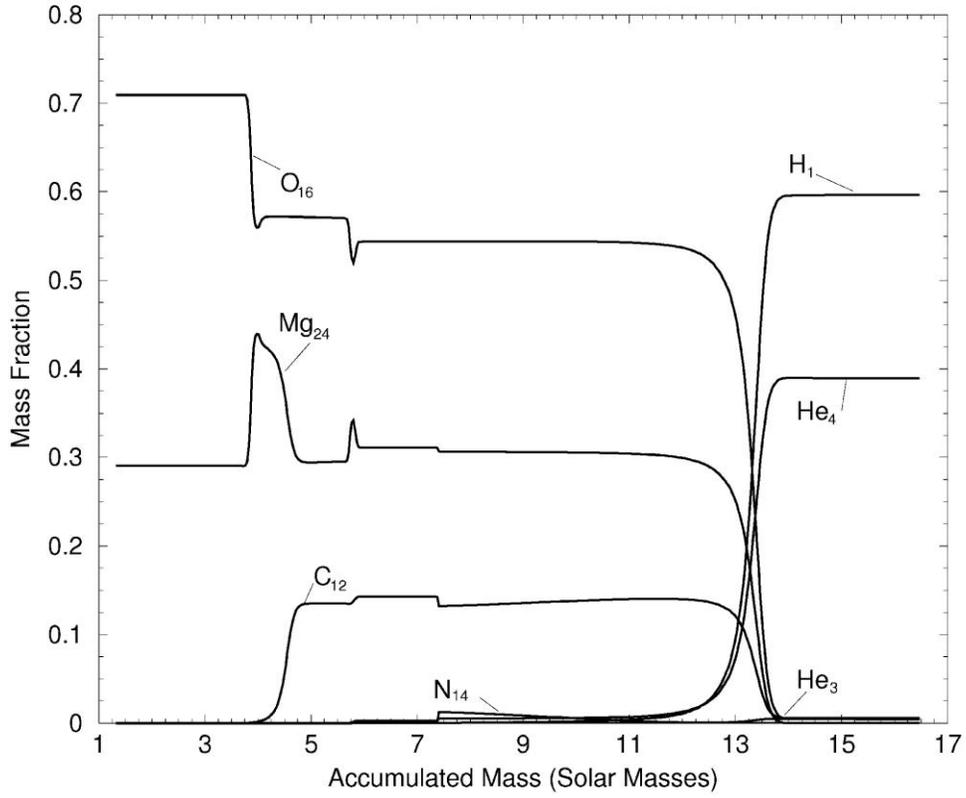

Figure 9. Mass Fractions (as labelled) against mass coordinate for the Pop III, 16.5 M$_\odot$ single star model. The O/Mg/C layer is 13 M$_\odot$, and a 3.5 M$_\odot$ H layer. The neutron star mass, 1.38 M$_\odot$, has been removed and set as an inner boundary. In this model, $^{24}$Mg represents all elements heavier than carbon.

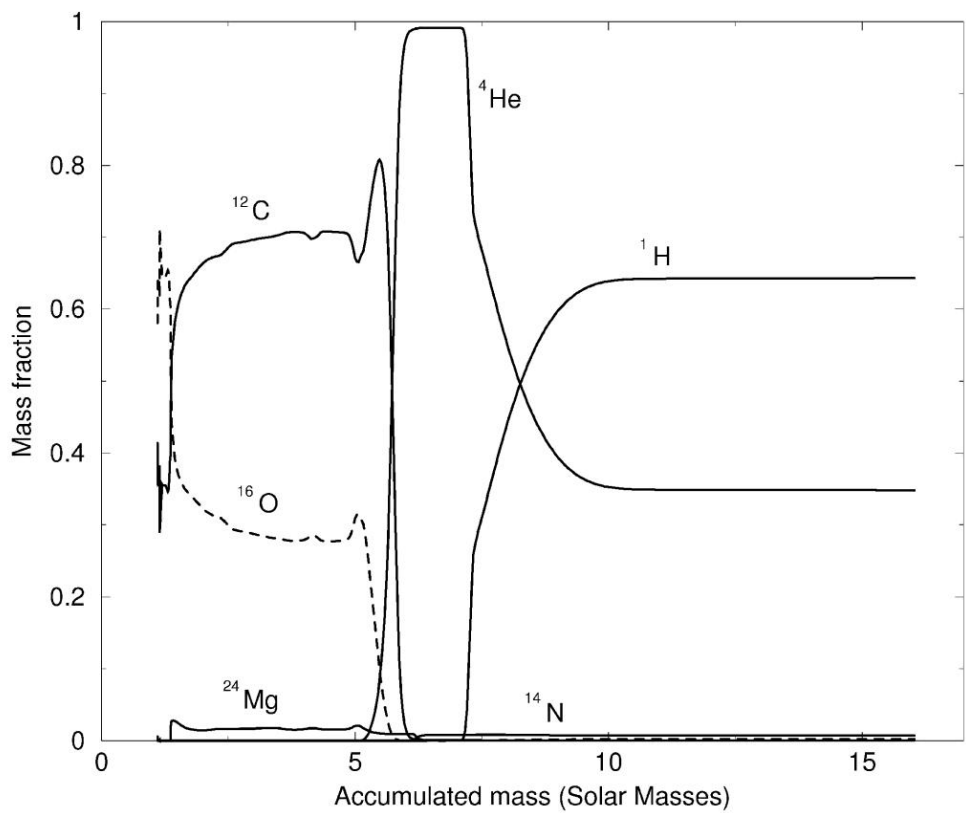

Figure 10. Mass Fractions (as labelled) against mass coordinate for the accretion model. The C/O layer is about 6 $M_\odot$, the He layer is 1.5 $M_\odot$, and the H layer is 8 $M_\odot$. The inner cut at 1.35 $M_\odot$ is the mass of the neutron star that is set as an inner boundary. In this model, $^{24}$Mg represents all elements heavier than carbon.

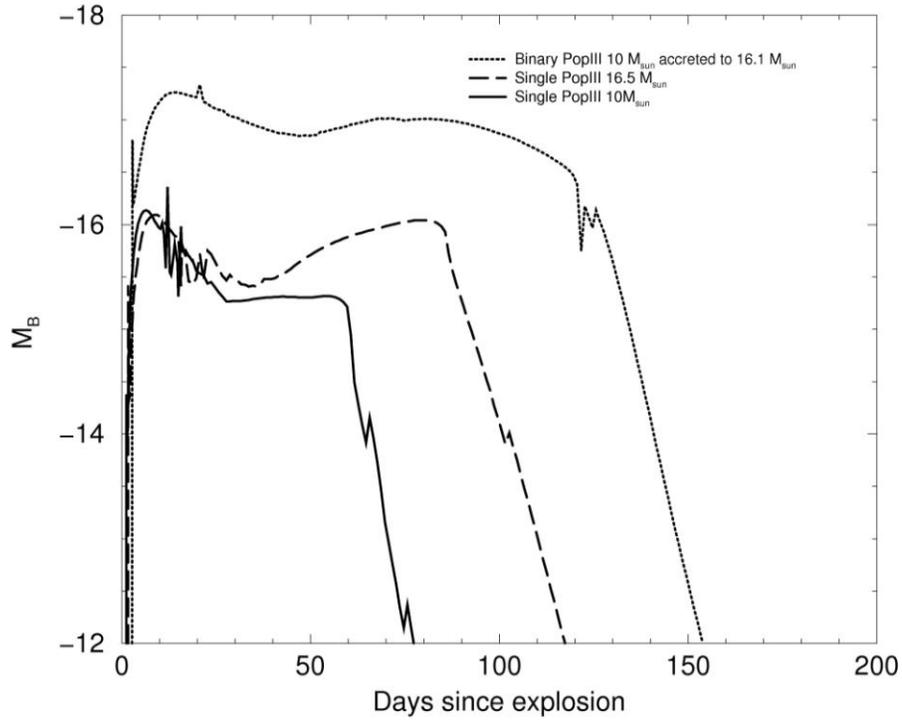

Figure 11. Pop III supernova light curves. The solid line is the 10 $M_\odot$ single star, the dashed line is the 16.5 $M_\odot$ single star, and the dotted line is the accretion model. The brightest model is the accretion model. There are two reasons for this: the radius is intrinsically larger than the other models (see Table 5) and the explosion energy needed was larger. The faintest light curve is the Pop III, 10 $M_\odot$ single star due to its small mass, which also gives a shorter plateau.

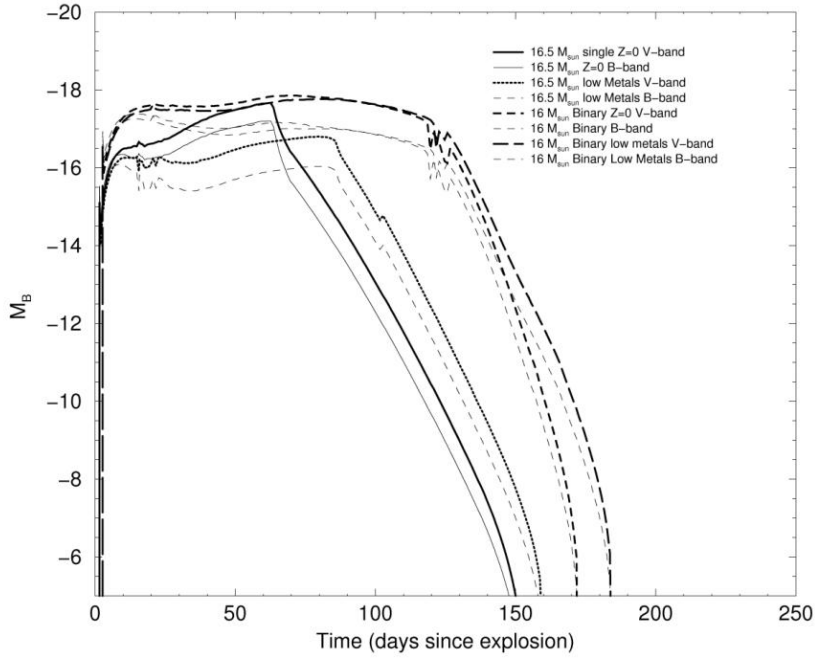

Figure 12. *B* and *V* band light curve comparison of similar mass models. The accretion model ends up with a mass of about 16.1 M$_\odot$. This is compared to the light curve for a single star of 16.5 M$_\odot$. The differences are significant mainly in the length of the plateau, being 60-80 days for the single star compared to 120 days for the binary model. The effect of changing the heavy metal abundance has little effect in the light curves of the binary model. However a definite change in the light curve is seen for the single star model. The peak is fainter and the plateau is longer for the higher heavy metal abundance progenitors.

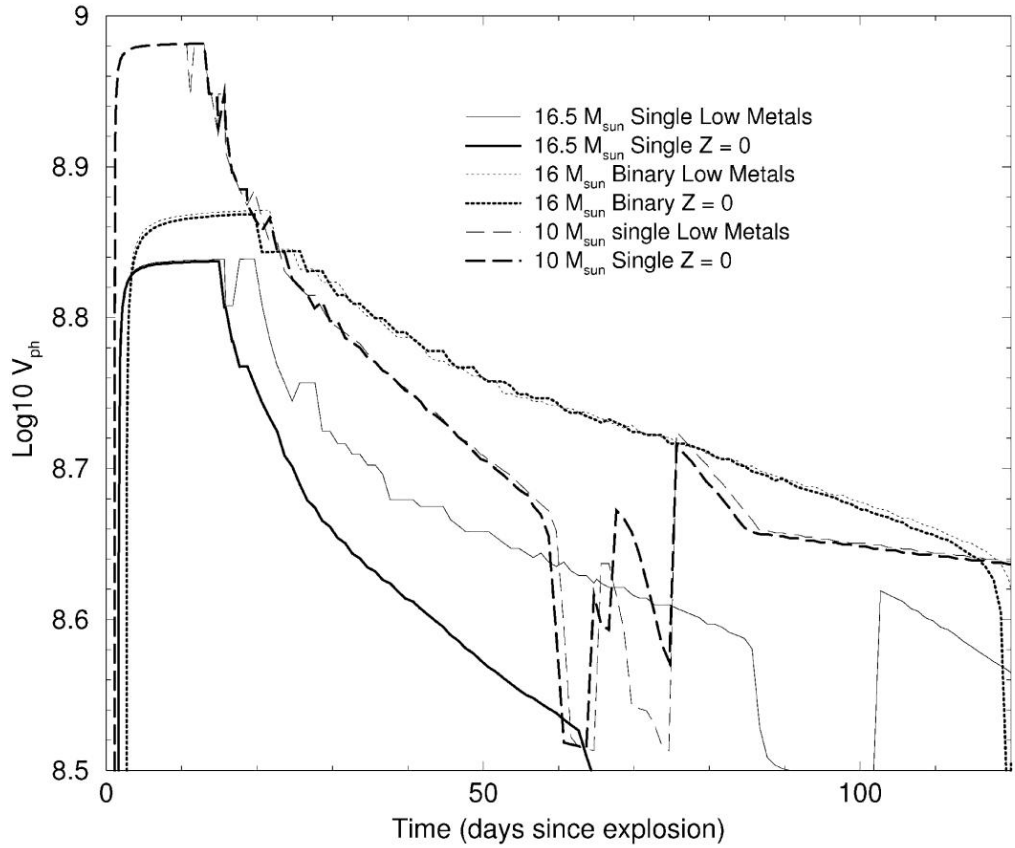

Figure 13. Photosphere velocity versus time. Three explosion models are compared each with two different metallicities of Z = 0.0 and Z = 0.008. The solid curve is 10 M$_\odot$ single star. The dotted curve is the accretion model, and the dashed curve is the 16.5 M$_\odot$ single star. Only the 16.5 M$_\odot$ model showed a difference in the photospheric velocity for different metallicities.

Table 1 Time-scales for core burning phases in single star evolution

| | Mass ($M_\odot$) | Z | $\tau_{MS}$ ($10^7$ yr) | $\tau_{HeCB}$ ($10^6$ yr) | $\tau_{CCB}$ ($10^4$ yr) |
|---|---|---|---|---|---|
| Sch02* | 9.0 | 0.00 | 2.02 | - | - |
| G-PSFG-B05* | 9.0 | 0.00 | 2.19 | 3.49 | 2.53 |
| *This work* | 10.0 | 0.00 | 1.75 | 2.03 | 3.30 |
| TSV02 | 10.0 | 0.00 | 1.76 | - | - |
| SLL02 | 10.0 | 0.00 | 1.82 | 3.00 | - |
| MGCW01* | 10.0 | 0.00 | 1.89 | 1.41 | - |
| *This work* | 10.0 | 0.02 | 2.08 | 3.87 | 1.27 |
| G-BI94 & RG-BI96* | 10.0 | 0.02 | 1.82 | 3.67 | 1.90 |
| IRG-B97* | 10.5 | 0.02 | 1.71 | 3.18 | 1.62 |
| *This work* | 16.5 | 0.00 | 0.96 | 0.83 | 0.82 |
| TSV02 | 15.0 | 0.00 | 1.11 | - | - |
| SLL02 | 15.0 | 0.00 | 1.21 | 0.90 | - |
| Sch02* | 15.0 | 0.00 | 1.04 | - | - |
| MGCW01** | 15.0 | 0.00 | 1.13 | 0.78 | - |
| TSV02 | 20.0 | 0.00 | 0.85 | - | - |
| SLL02 | 20.0 | 0.00 | 0.88 | 0.63 | - |
| Sch02* | 25.0 | 0.00 | 0.65 | - | - |
| MGCW01* | 20.0 | 0.00 | 0.30 | 0.58 | |

*Constant mass models. G-BI94 = García-Berro & Iben (1994), G-PSFG-B05 = Gil-Pons et al. (2005), IRG-B97 = Iben et al. (1997), MGCW01 = Marigo et al. (2001), RG-BI96 = Ritossa et al. (1996), Sch02 = Schaerer (2002), T SLL02 = Siess, Livio & Lattanzio (2002), SV02 = Tumlinson, Shull & Venkatesan (2003).

Table 2. Selected physical characteristics of Pop III stars

| M=10 M$_\odot$<br>Z = 0.00 | Onset of H-<br>core burning | Onset of He-<br>core burning | Onset of C-<br>core burning | Last computed<br>model |
|---|---|---|---|---|
| M interior to H-shell (M$_\odot$) | 0.00 | 2.08 | 2.25 | 1.36 |
| Log(L/L$_\odot$) | 3.92 | 4.19 | 4.23 | 5.18 |
| Log(T$_{eff}$) | 4.37 | 4.61 | 4.23 | 3.58 |
| Log(R)(cm) | 11.60 | 11.25 | 12.03 | 13.79 |
| dM/dt (M$_\odot$/yr) | 2.26 $10^{-15}$ | 8.10 $10^{-15}$ | 9.40 $10^{-16}$ | 2.0 $10^{-6}$ |
| M$_H$ (M$_\odot$) | 7.65 | 5.65 | 5.37 | 5.44 |
| M$_{He}$ (M$_\odot$) | 2.35 | 4.35 | 3.27 | 3.17 |
| Log(T$_c$) | 7.29 | 8.13 | 8.72 | 8.52 |
| Log($\rho_c$) | 0.54 | 3.39 | 5.93 | 8.65 |
| $\nu_{loss}$ | 6.45 $10^{-7}$ | 0.51 | 1.42 $10^4$ | 3.9 $10^5$ |
| $\varepsilon_g$ | 8.30 $10^3$ | -1.43 $10^3$ | 5.47 $10^3$ | 4.29 $10^5$ |
| Y$_c$ | 0.235 | 0.996 | 2.00 $10^{-13}$ | 1.14 $10^{-14}$ |

| M=16.5 M$_\odot$<br>Z = 0.00 | Onset of H-<br>core burning | Onset of He-<br>core burning | Onset of C<br>-core burning | Last computed<br>model |
|---|---|---|---|---|
| M interior to H-shell (M$_\odot$) | 0.00 | 4.92 | 2.54 | 2.33 |
| Log(L/L$_\odot$) | 4.52 | 4.80 | 4.71 | 5.26 |
| Log(T$_{eff}$) | 4.43 | 4.68 | 3.97 | 3.49 |
| Log(R)(cm) | 11.77 | 11.42 | 12.77 | 14.00 |
| dM/dt (M$_\odot$/yr) | 2.2 x $10^{-14}$ | 8.60 x $10^{-14}$ | 2.29 x $10^{-8}$ | 1.10 x $10^{-4}$ |
| M$_H$ (M$_\odot$) | 12.62 | 8.58 | 8.33 | 8.33 |
| M$_{He}$ (M$_\odot$) | 3.88 | 7.92 | 5.44 | 3.68 |
| Log(T$_c$) | 7.32 | 8.16 | 8.76 | 8.60 |
| Log($\rho_c$) | 0.25 | 2.61 | 5.12 | 8.13 |
| $\nu_{loss}$ | 1.86 x $10^{-6}$ | 1.89 | 3.11 x $10^4$ | 1.96 x $10^6$ |
| $\varepsilon_g$ | 3.32 x $10^4$ | 4.61 x $10^4$ | 4.80 x $10^4$ | 1.87 x $10^6$ |
| Y$_c$ | 0.235 | 0.999 | 2.00 x $10^{-14}$ | 2.00 x $10^{-14}$ |

Table 3. Selected surface abundances in mass fractions

| M=10.0 M$_\odot$<br>Z = 0.00 | MS through He$_c$<br>burning | Following peak<br>C burning | Last computed<br>model |
|---|---|---|---|
| X | 0.765 | 0.643 | 0.630 |
| Y | 0.235 | 0.357 | 0.369 |
| C | 0.000 | 8.13 $10^{-7}$ | 3.87 $10^{-4}$ |
| N | 0.000 | 4.86 $10^{-10}$ | 5.50 $10^{-6}$ |
| O | 0.000 | 8.29 $10^{-9}$ | 1.07 $10^{-4}$ |
| Mg and other | 0.000 | 9.89 $10^{-17}$ | 7.01 $10^{-8}$ |

| M=16.5 M$_\odot$<br>Z = 0.00 | MS through He$_c$<br>burning | Following peak<br>C burning | Last computed<br>model |
|---|---|---|---|
| X | 0.765 | 0.716 | 0.597 |
| Y | 0.235 | 0.283 | 0.389 |
| C | 0.000 | 3.91 $10^{-4}$ | 5.27 $10^{-3}$ |
| N | 0.000 | 6.60 $10^{-4}$ | 4.53 $10^{-3}$ |
| O | 0.000 | 5.58 $10^{-5}$ | 4.14 $10^{-3}$ |
| Mg and other | 0.000 | 2.41 $10^{-9}$ | 1.52 $10^{-7}$ |

Table 4. Final surface abundances for our accretion model.

| 10M$_\odot$, Z = 0.00<br>M$_{final}$ = 16.1 M$_\odot$ | Final surface abundance (mass fractions) |
|---|---|
| X | 0.649 |
| Y | 0.343 |
| C | 1.20 10$^{-4}$ |
| N | 6.46 10$^{-3}$ |
| O | 1.81 10$^{-3}$ |
| Mg and other | 1.03 10$^{-8}$ |

Table 5. Model parameters given by the evolutionary code and the parameters used in the explosion simulation.

| Model | Pre-explosion Mass (M$_\odot$) | Pre-explosion Radius (R$_\odot$) | Explosion Energy (foe) | Ejected Mass (M$_\odot$) | $^{56}$Ni Mixing thru He-core (M$_\odot$) | Composition |
|---|---|---|---|---|---|---|
| Single | 10 | 270.8 | 1.16 | 8.9 | 2.62 | Z = 0 & Z = 0.008 |
| Single | 16 | 457.1 | 1.14 | 15.1 | 10.0 | Z = 0 & Z = 0.008 |
| Accretion | 16.04 | 1262.9 | 1.90 | 15.0 | 9.0 | Z = 0 & Z = 0.008 |